\documentclass[12pt]{article}
\usepackage{amssymb,amsmath,epsfig}

\begin{document}

\title{\bf Stability of Accelerated Expansion in Nonlinear Electrodynamics}
\author{M. Sharif \thanks {msharif.math@pu.edu.pk} and Saadia Mumtaz
\thanks{sadiamumtaz17@gmail.com}\\
Department of Mathematics, University of the Punjab,\\
Quaid-e-Azam Campus, Lahore-54590, Pakistan.}

\date{}
\maketitle

\begin{abstract}
This paper is devoted to study the phase space analysis of isotropic
and homogenous universe model by taking a noninteracting mixture of
electromagnetic and viscous radiating fluids whose viscous pressure
satisfies a nonlinear version of the Israel-Stewart transport
equation. We establish an autonomous system of equations by
introducing normalized dimensionless variables. In order to analyze
stability of the system, we find corresponding critical points for
different values of the parameters. We also evaluate power-law scale
factor whose behavior indicates different phases of the universe
model. It is concluded that bulk viscosity as well as
electromagnetic field enhances the stability of accelerated
expansion of the isotropic and homogeneous universe model.
\end{abstract}

\section{Introduction}

Many astronomical observations (type Ia supernova, large scale
structure and cosmic microwave background radiation) predict that
our universe is expanding at an accelerating rate in its present
stage \cite{1}. These observations suggest two cosmos phases, i.e.,
the cosmic state before radiation (the primordial inflationary era)
and ultimately the present cosmos phase after the matter dominated
era. In the last couple of decades, it is speculated that the source
for this observed cosmic acceleration with unusual
anti-gravitational force may be an anonymous energy component dubbed
as dark energy (DE). The existence of this energy with large
negative pressure can be recognized by its distinctive nature from
ordinary matter which may lead to cosmic expansion. The study of the
dominant contents of matter distribution in the universe has
remained one of the most challenging issues. Recent observations
show that the visible part of our universe is made up of baryonic
matter contributing only $5\%$ of the total budget while the
remaining ingredients yield the total energy density composed of
non-baryonic fluids ($68\%$ DE and $27\%$ dark matter) \cite{2}.

Several cosmological proposals have been introduced in literature to
explore the ambiguous nature of DE. The cosmological constant
($\Lambda$) governed by a negative equation of state (EoS) parameter
($\gamma=-1$) is taken to be the simplest characterization of DE.
However, this identification has two well-known problems, i.e.,
fine-tuning and cosmic coincidence. In addition, there are several
cosmological models which can be considered as an alternative to
$\Lambda$ like scalar field model \cite{5}, phantom model \cite{6},
tachyon field \cite{7} and k-essence \cite{8} that also suggest
expanding behavior of the universe. Another approach involves the
generalization of simple barotropic EoS to more exotic forms such as
Chaplygin gas \cite{9} and its modification \cite{10}. It has also
been demonstrated that a fluid with bulk viscosity may cause
accelerated expansion of the universe model without cosmological
constant or scalar field \cite{10a}. Our main concern is to find
another approach which can minimize exotic forms of matter by
introducing dissipation through viscous effects of fluids.

During the last few years, cosmological models including nonlinear
electromagnetic fields have attained remarkable interest. The
application of this electrodynamics to different universe models may
lead to many significant results. Nonlinear electrodynamics (NLED)
is the generalization of Maxwell theory which is considered as the
most viable theory to remove the initial singularities. Vollick
\cite{10b} considered FRW universe model with NLED and found that
the respective model will show a period of late-time acceleration
for $E^2<3B^2$. Kruglov \cite{10c} found that the universe tends to
accelerate in magnetic background at the early era due to NLED
model. Ovgun \cite{10d} formulated analytical nonsingular extension
of isotropic and homogeneous solutions by presenting a new
mathematical model in nonlinear magnetic monopole fields.

The study of possible stable late-time attractors has attained
remarkable significance for different universe models. A phase space
analysis manifests dynamical behavior of a cosmological model
through a global view by reducing complexity of the equations
(converting the system of equations to an autonomous system) which
may help to understand different stages of evolution. Copeland et
al. \cite{12} studied a phase plane analysis of standard
inflationary models and found that these models cannot solve density
problem. Guo et al. \cite{13} explored phase space analysis of FRW
universe model filled with barotropic fluid and phantom scalar field
in which phantom dominated solution is found to be a stable
late-time attractor.

Garcia-Salcedo \cite{13a} examined the dynamics of FRW universe with
NLED and found that the critical points have no effects. Yang and
Gao \cite{14} discussed phase space analysis of k-essence cosmology
in which critical points play an important role for the final state
of the universe. Xiao and Zhu \cite{15} analyzed stability of FRW
universe model in loop quantum gravity via phase space portraits by
taking barotropic fluid as well as positive field potential.
Acquaviva and Beesham \cite{16} studied phase space analysis of FRW
spacetime filled with noninteracting mixture of fluids (dust and
viscous radiation) and found that nonlinear viscous model shows the
possibility of current cosmic expansion.

This paper is devoted to study the phase space analysis of FRW
universe model with nonlinear viscous fluid. The plan of the paper
is as follows. In section \textbf{2}, we provide basic formalism for
NLED and general equations as well as a nonlinear model for bulk
viscosity. An autonomous system of equations is established to
analyze stability of the system by introducing normalized
dimensionless variables in section \textbf{3}. Section \textbf{4}
provides the formulation of power-law scale factor. Finally, we
conclude our results in the last section.

\section{Nonlinear Electrodynamics and General Equations}

The standard cosmological model is successful in resolving many
issues but still there are some issues which remain to be solved.
One of them the initial singularity which leads to a troubling state
of affairs because at this point, all known physical theories break
down. If the early universe is governed by Maxwell's equations, then
there will be a spacelike initial singularity in the past. However,
if Maxwell's equations become modified in the early universe, when
the electromagnetic field is large, it might help avoiding the
occurrence of cosmic singularities \cite{aa}. For the situations
where strong electromagnetic field occurs, it makes sense to couple
gravitation with NLED. The coupling of Einstein gravity with NLED is
defined by the action
\begin{equation}
S=\frac{1}{16\pi}\int\sqrt{-g}[R-\mathcal{L}(F,F^*)]d^4x.
\end{equation}
We consider nonlinear extension of Maxwell Lagrangian density up to
second order terms in the field invariants $F=F_{\mu\nu}F^{\mu\nu}$
and $F^*=F^*_{\mu\nu}F^{\mu\nu}$ given by \cite{10b}
\begin{equation}\label{a}
\mathcal{L}=\mathcal{L}(F,F^*)=-\frac{1}{4\mu_{0}}F+\alpha F^2+\beta
F^{*2},
\end{equation}
where $\mu_{0}$ denotes magnetic permeability, $\alpha,\beta>0$ are
arbitrary constants which yield linear density for
$\alpha,\beta\rightarrow0$ and $F^*_{\mu\nu}$ is the dual of
electromagnetic field tensor. We do not consider the term $FF^*$
involving $F^*$ in order to preserve the parity \cite{10bb, 10bbb}.
The linear term of this Lagrangian dominates during radiation
dominated era while the quadratic terms dominate in the early
universe that corresponds to the bouncing behavior of the universe
to avoid initial singularity \cite{bb}. The mechanisms behind the
bounce have been demonstrated in \cite{10b1,10b2}. The
energy-momentum tensor associated with this Lagrangian has the
following form
\begin{equation}\label{b}
T_{\mu\nu(EM)}=-4\partial_{F}\mathcal{L}F^{\eta}_{\mu}F_{\eta\nu}+
(\partial_{F^*}\mathcal{L}F^*-\mathcal{L})g_{\mu\nu}.
\end{equation}
In order to fulfill the requirement of isotropic and homogeneous
universe, i.e., the electromagnetic field to act as its source, the
energy density and the pressure corresponding to the electromagnetic
field can be computed by averaging over volume \cite{10bbb, 10b1}.
It is assumed that electric and magnetic fields have coherent
lengths that are much shorter than the cosmological horizon scales.
After several conditions, the energy momentum tensor of the
electromagnetic field associated with $\mathcal{L}(F,F^*)$ can be
written as that of a perfect fluid
\begin{equation}
T_{\mu\nu}=(\rho+p)u_{\mu}u_{\nu}+pg_{\mu\nu},
\end{equation}
such that
\begin{eqnarray}\label{1a}
\rho_{EM}&=&-\mathcal{L}-4E^2\partial_{F}\mathcal{L},
\\\label{1b}
p_{EM}&=&\mathcal{L}-\frac{4}{3}(2B^2-E^2)\partial_{F}\mathcal{L},
\end{eqnarray}
where $\partial_{F}$ represents partial derivative with respect to
$F=F_{\mu\nu}F^{\mu\nu}=2(B^2-E^2)$, $E$ and $B$ denote the averaged
electric and magnetic fields, respectively.

We consider isotropic and homogeneous universe model given by
\begin{equation}\label{1}
ds^2=-dt^2+a(t)(dr^2+r^2d\theta^2+r^2\sin^2\theta d\phi^2),
\end{equation}
where $a(t)$ is the scale factor. We assume the universe model to be
filled with two cosmic fluids, i.e., a noninteracting
electromagnetic fluid with energy density $\rho_{EM}$ as well as
pressure $p_{EM}$ and a viscous fluid having energy density
$\rho_{v}$ as well as pressure $p=p_{v}(\rho_{v})+\Psi$. Here
$p_{v}$ represents the equilibrium part of viscous pressure whereas
$\Psi$ is the non-equilibrium part, i.e., bulk viscous pressure
satisfying an evolution equation. Bulk viscosity plays an important
role to stabilize the density evolution and overcomes the rapid
changes in cosmos. It also promotes negative energy field in the
fluid and hence can play the role of dark energy to describe the
dynamics of cosmos. It has been suggested that a fluid with bulk
viscosity may cause an accelerated expansion of the universe model
without cosmological constant or scalar field \cite{10bb1}. The main
contribution of bulk viscosity to the effective pressure is its
dissipative effect. We obtain Raychaudhuri and constraint equations
from the field equations given by
\begin{eqnarray}\label{2}
\dot{\Theta}&=&-\frac{1}{3}\Theta^2-\frac{1}{2}\left[\rho_{EM}+
\rho_{v}+3(p_{EM}+p_{v}+\Psi)\right],\\\label{3}
0&=&\rho_{EM}+\rho_{v}-\frac{1}{3}\Theta^2,
\end{eqnarray}
where dot means derivative with respect to time. The conservation of
energy-momentum tensor yields the following evolution equations for
viscous and electromagnetic field components
\begin{eqnarray}\label{4}
\dot{\rho}_{v}&=&-[\rho_{v}+p_{v}+\Psi]\Theta, \\\label{5}
\dot{\rho}_{EM}&=&-[\rho_{EM}+p_{EM}]\Theta.
\end{eqnarray}

We consider a barotropic EoS for viscous fluid defined by
\begin{equation}\label{6}
p_{v}=(\gamma-1)\rho_{v},
\end{equation}
where $1\leq \gamma\leq2$. Using Eqs.(\ref{2}) and (\ref{3}),
Raychaudhuri and conservation equations for viscous fluid turn out
to be
\begin{eqnarray}\label{7}
\dot{\Theta}&=&-\frac{1}{2}\Theta^2-\frac{3}{2}[p_{EM}+(\gamma-1)
\rho_{v}+\Psi],\\\label{8}
\dot{\rho}_{v}&=&-[\gamma\rho_{v}+\Psi]\Theta.
\end{eqnarray}
We characterize the viscous pressure variable by the following
evolution equation \cite{b}
\begin{equation}\label{9}
\tau\dot{\Psi}=-\zeta\Theta-\Psi\left(1+\frac{\tau_{*}}
{\zeta}\Psi\right)^{-1}-\frac{1}{2}\tau\Psi\left[\Theta+\frac{\dot
{\tau}}{\tau}-\frac{\dot{\zeta}}{\zeta}-\frac{\dot{T}}{T}\right],
\end{equation}
where $\zeta$, $T$, $\tau$ and $\tau_{*}$ denote bulk viscosity,
local equilibrium temperature, linear relaxation time and
characteristic time in nonlinear background, respectively. This
equation is derived by using a nonlinear model describing a
relationship between thermodynamic flux ``$\Psi$'' and thermodynamic
force ``$\chi$" in the form
\begin{equation}\label{10}
\Psi=-\frac{\zeta\chi}{1+\tau_{*}\chi}.
\end{equation}
This is a nonlinear extension of Israel-Stewart equation which
reduces to its linear form as $\tau_{*}\rightarrow0$. The nonlinear
term in Eq.(\ref{9}) must be positive for thermodynamic consistency
and positivity of entropy production rate. The parameters involved
in Eq.(\ref{9}) can be defined by the relations
$\zeta=\zeta_{0}\Theta$ $(\zeta>0)$, $\tau=\frac{\zeta}{\gamma
\nu^2\rho_{v}}$, $\tau_{*}=k^2\tau$ and
$T=T_{0}\rho^{(\gamma-1)/\gamma}$. Here $k$ is a constant such that
$k=0$ gives linear (Israel-Steward) case while $T_{0}$ represents
constant temperature. Also, $\nu$ corresponds to the dissipative
effect of the speed of sound $V$ such that $V^2=c^2_{s}+\nu^2$,
where $c^2_{s}$ is its adiabatic contribution. By causality,
$V\leq1$ and $c^2_{s}=\gamma-1$ which yields
\begin{equation}\label{11}
\nu^2\leq2-\gamma, \quad 1\leq \gamma\leq2.
\end{equation}
The explicit form of evolution equation by using the above relations
yields
\begin{equation}\label{12}
\dot{\Psi}=-\gamma\nu^2\rho_{v}\Theta-\frac{\gamma
\nu^2\Psi\rho_{v}}{\zeta_{0}\Theta} \left(1+\frac{k^2\Psi}{\gamma
\nu^2\rho_{v}}\right)^{-1}-\frac{1}{2}
\Psi\left[\Theta-\left(\frac{2\gamma-1}{\gamma}\right)
\frac{\dot{\rho_{v}}}{\rho_{v}}\right].
\end{equation}

\section{Phase Space Analysis}

In this section, we discuss the phase space analysis of isotropic
and homogeneous universe model for radiation case. Due to many
arbitrary parameters, it seems difficult to find analytical solution
of the evolution equation. In this context, we define normalized
dimensionless variables $\Omega=\frac{3\rho_{v}}{\Theta^2}$ and
$\tilde{\Psi}=\frac{3\Psi}{\Theta^2}$ such that the corresponding
dynamical system can be reduced to autonomous one. We also define a
new variable $\tilde{\tau}$ for time through which the corresponding
derivative is represented by prime such that
$\frac{dt}{d\tilde{\tau}}=\frac{3}{\Theta}$. Here each term is
associated with some physical explicit background since the chosen
dimensionless variables $\Omega$ and $\tilde{\Psi}$ occur due to
physical impact of viscous energy density and pressure,
respectively. The system of Eqs.(\ref{7}) and (\ref{8}) in terms of
these normalized variables takes the form
\begin{eqnarray}\label{13}
\frac{\Theta'}{\Theta}&=&-\frac{3}{2}\left[1+p_{EM}+(\gamma-1)\Omega+
\tilde{\Psi}\right],\\\label{14}\frac{3\rho_{v}'}{\Theta^2}&=&-
3[\gamma\Omega+\tilde{\Psi}].
\end{eqnarray}
Differentiation of the dimensionless variable for energy density
gives
\begin{equation}\label{15}
\Omega'=\frac{3\rho_{v}'}{\Theta^2} -2\Omega\frac{\Theta'}{\Theta}.
\end{equation}
Using Eqs.(\ref{13}) and (\ref{14}), this equation turns out to be
\begin{equation}\label{16}
\Omega'=3(\Omega-1)[\Omega(\gamma-1)+\tilde{\Psi}+3p_{EM}].
\end{equation}
Now we introduce the concept of a new evolution equation for
$\tilde{\Psi}$. The first derivative of $\tilde{\Psi}$ with respect
to $\tilde{\tau}$ through Eq.(\ref{13}) leads to an evolution
equation of the form
\begin{eqnarray}\nonumber
\tilde{\Psi}'&=&-3\gamma \nu^2\Omega\left[1+\frac{\tilde{\Psi}}
{3\zeta_{0}}\left(1+\frac{k^2\tilde{\Psi}}{\gamma\nu^2\Omega}
\right)^{-1}\right]\\\label{17}&+&3\tilde{\Psi}\left[1+3p_{EM}
\left(1-\frac{3}{\Omega}\frac{2\gamma-1}{2\gamma}\right)\right]
-3(\gamma-1)\tilde{\Psi}(1-\Omega).
\end{eqnarray}
It is mentioned here that Eqs.(\ref{16}) and (\ref{17}) play a
remarkable role to describe the respective dynamical system for
phase space analysis.

In order to find the critical points
$\{\Omega_{c},\tilde{\Psi}_{c}\}$, we need to solve the dynamical
system by imposing the condition $\Omega'=\tilde{\Psi}'=0$. The
stability of FRW universe model can be analyzed according to the
nature of critical points. Here we restrict the phase space region
to a condition which is necessary for the positivity of entropy
production rate given by \cite{16,b}
\begin{equation}\label{18}
\tilde{\Psi}>-\frac{\gamma\nu^2\Omega}{k^2}.
\end{equation}
This condition tends the possible negative values of $\tilde{\Psi}$
towards zero for $k^2\gg\nu^2$. Contrarily, the bulk pressure will
be less restrictive if $k^2\ll \nu^2$. It is noted that finite
values of $k$ allow only positive values of bulk pressure in the
limit $\nu\rightarrow0$. It would be more convenient to consider
$k^2\leq\nu^2$ along with $\nu^2\leq2-\gamma$ and $\tau_{*}=k^2\tau$
which leads to the fact that the characteristic time for nonlinear
effects $\tau_{*}$ does not exceed the characteristic time in linear
background $\tau$. We characterize the critical points by
deceleration parameter $q=-1-\frac{\Theta'}{\Theta}$ and effective
EoS parameter $\gamma_{eff}=-\frac{2\Theta'}{3\Theta}$ which yield
\begin{eqnarray}\label{19}
q&=&\frac{1}{2}\left[1+9p_{EM}+3(\gamma-1)\Omega+3\tilde{\Psi}\right],
\\\label{20}
\gamma_{eff}&=&1+3p_{EM}+(\gamma-1)\Omega+\tilde{\Psi}.
\end{eqnarray}
To examine a region of phase space undergoing accelerated expansion,
we impose $q<0$ in Eq.(\ref{19}) which gives
\begin{equation}\label{c}
\tilde{\Psi}<-\frac{1}{3}-3p_{EM}-(\gamma-1)\Omega.
\end{equation}
The possibility of accelerated expansion in the physical phase space
is determined by comparing Eqs.(\ref{18}) and (\ref{19}) through
$q<0$ given by
\begin{equation}\label{21a}
\frac{\nu^2}{k^2}>\frac{1+9p_{EM}+3(\gamma-1)\Omega}{3\gamma\Omega}.
\end{equation}
Substituting $\Omega'=0$ in Eq.(\ref{16}), we find the following
conditions
\begin{eqnarray}\label{21}
\Omega_{c}&=&1,\\\label{22}
(\gamma-1)\Omega_{c}+\tilde{\Psi}_{c}+3p_{EM}&=&0.
\end{eqnarray}
We insert these conditions in $\tilde{\Psi}'$ to find the location
of critical points. This analysis is carried out by characterizing
the viscous fluid through the choice of its EoS parameter $\gamma$
(radiation). We consider $0<k^2=\nu^2\leq2-\gamma$ for which the
case of stiff matter ($\gamma=2$) is excluded from the analysis as
it yields $\nu^2=0$.

\subsection{Radiation Case $(\gamma=\frac{4}{3})$}

We consider the radiation case for phase space analysis  by taking
$\gamma=\frac{4}{3}$. Imposing the condition (\ref{21}) and
$\tilde{\Psi}'=0$ in Eq.(\ref{17}), we have
\begin{eqnarray}\label{27}
&&\frac{3\nu^2}{4\zeta_{0}}\tilde{\Psi}^3-
\frac{\nu^2}{\zeta_{0}}\tilde{\Psi}^2-3\tilde{\Psi}\left(1-\frac{4\nu^2}
{9\zeta_{0}}-\frac{21p_{EM}}{8}\right)+4\nu^2=0.
\end{eqnarray}
This cubic equation yields three roots among which we retain only
those roots that lie in the physical phase space. The general form
of the dynamical system is given by
\begin{equation}\label{24}
\Omega'=X(\Omega, \tilde{\Psi}),\quad \tilde{\Psi}'=Y(\Omega,
\tilde{\Psi}).
\end{equation}
The eigenvalues of the system can be determined by the Jacobian
matrix
\begin{equation}
Z=\left(
\begin{array}{cc}
\frac{\partial X}{\partial\Omega}&\frac{\partial X}{\partial\tilde{\Psi}}\\
\frac{\partial Y}{\partial\Omega}&\frac{\partial Y}{\partial\tilde{\Psi}}\\
\end{array}
\right)_{|P_i^{\pm}}.
\end{equation}
The eigenvalues for the above stability matrix corresponding to the
points $P^{\pm}_{r}$ are given by
\begin{eqnarray}\label{28a}
\lambda_{1}&=&1+3\tilde{\Psi}+9p_{EM},
\\\label{28b}\lambda_{2}&=&-\frac{16\nu^2}{3\zeta_{0}}\left[\frac{1}{4
+\tilde{\Psi}}-\frac{\tilde{\Psi}}{(4+\tilde{\Psi})^2}\right]
-\frac{63p_{EM}}{8}+6\tilde{\Psi}+3.
\end{eqnarray}
The fixed point is called a source (respectively, a sink) if both
eigenvalues consist of positive (respectively, negative) real parts.
In case of viscous radiating fluid, we can explore source and sink
according to the sign of eigenvalues as well as direction of the
trajectories. We investigate two critical points $P_{r}^+=\{1,
\tilde{\Psi}_{c}^+\}$ and $P_{r}^-=\{1, \tilde{\Psi}_{c}^-\}$
corresponding to positive $(\tilde{\Psi}_{c}^+)$ and negative
$(\tilde{\Psi}_{c}^-)$ roots, respectively. By taking $\Omega_{c}=0$
and the second condition (\ref{22}) with
$\tilde{\Psi}_{c}=-\frac{\Omega_{c}}{3}-3p_{EM}$, we obtain
$P^0_{r}=\{0,-3p_{EM}\}$.

\subsubsection{Case I:}

We are interested to analyze the impact of electromagnetic field on
stability of the critical points in the presence of nonlinear bulk
viscosity. The energy density (\ref{1a}) and pressure (\ref{1b}) are
given by
\begin{eqnarray}\label{ss}
\rho_{EM}&=&\frac{1}{2\mu_{0}}(B^2+E^2)-4\alpha(B^2-E^2)(B^2+3E^2),
\\\label{ss1}
p_{EM}&=&\frac{1}{6\mu_{0}}(B^2+E^2)-\frac{4\alpha}{3}(B^2-E^2)(5B^2-E^2).
\end{eqnarray}
The dynamical behavior of critical points for different values of
electric and magnetic fields as well as other parameters is shown in
Figures \textbf{1-2}. The green trajectory depicts a flow from the
point $P^{+}_{d}$ towards $P^{-}_{d}$. The white region corresponds
to the negative entropy production rate that diverges on its
boundary whereas the green region shows accelerated expansion of the
universe ($q<0$). Here the point $P^{0}_{d}$ shows varying behavior,
i.e., either it is a saddle point or a sink depending on the values
of different parameters.

In these plots, we have taken $\zeta_{0}=0.2,1$ by varying $\nu,k,B$
and $E$. For $\nu=k=\sqrt{1/5}$ and $\zeta_{0}=0.2$, it is found
that the global attractor $P_{d}^-$ lies in green region showing
accelerated expansion for the same values of $B$ and $E$. This
region tends to decrease by increasing $E$ such that the point
$P_{d}^-$ lies in the deceleration region. By increasing
$\zeta_{0}$, we find accelerated expansion with different values of
$B,E$ and larger values of the parameters $\nu$ and $k$. For
$\nu=k=1$ and $\zeta_{0}=1$, we find accelerated expansion of the
universe model for all choices of electric and magnetic fields. The
point $P^{0}_{d}$ behaves as a sink for $\zeta_{0}=0.2$ which
becomes a saddle point for larger values of $\zeta_{0}$. We observe
that the increasing value of bulk viscosity increases the region for
accelerated expansion in the presence of NLED. In the following, we
discuss two different cases for electric as well as magnetic
universe.
\begin{figure}\center
\epsfig{file=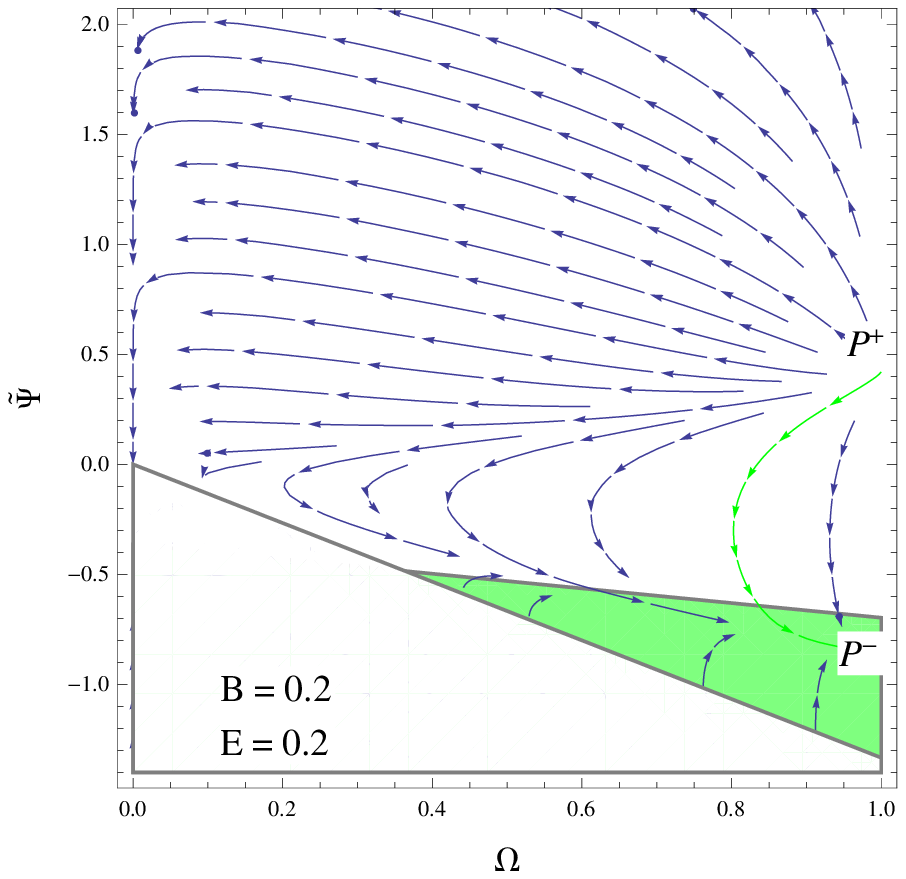,width=0.4\linewidth}\epsfig{file=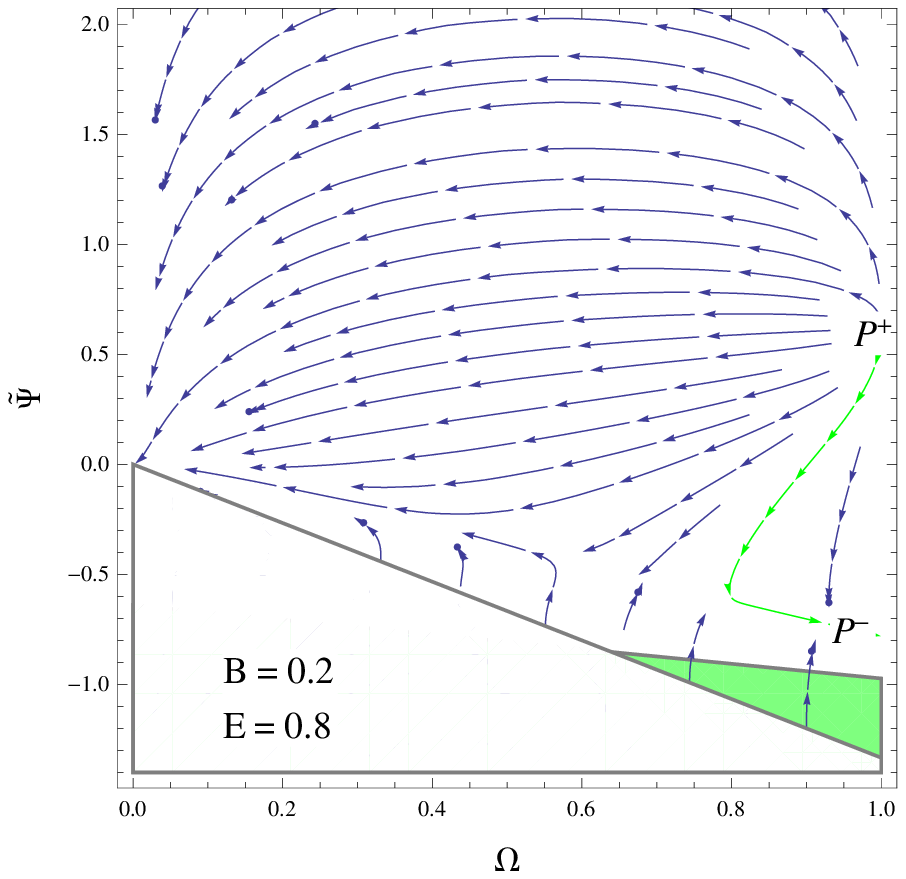,
width=0.4\linewidth}\caption{Plots for the phase plane evolution of
viscous radiating fluid with $\gamma=4/3$, $\nu=k=\sqrt{1/5}$,
$\zeta_{0}=0.2$, $\alpha=0.01$ and different values of $B$ and $E$.}
\end{figure}
\begin{figure}\center
\epsfig{file=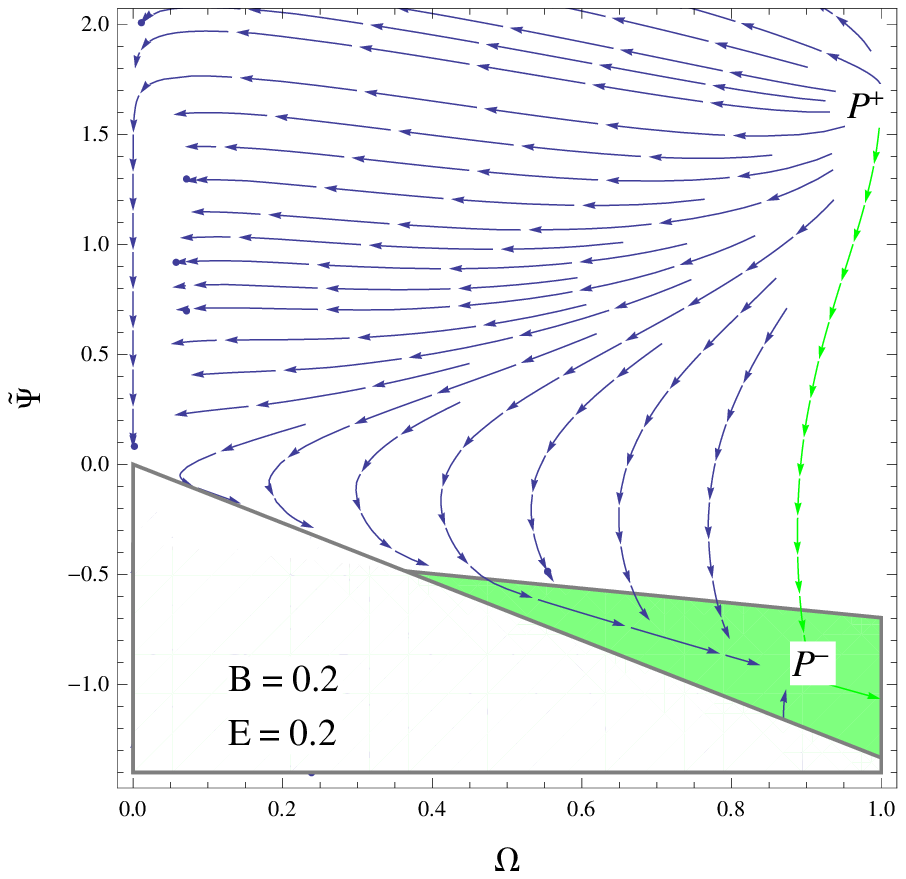,width=0.4\linewidth}\epsfig{file=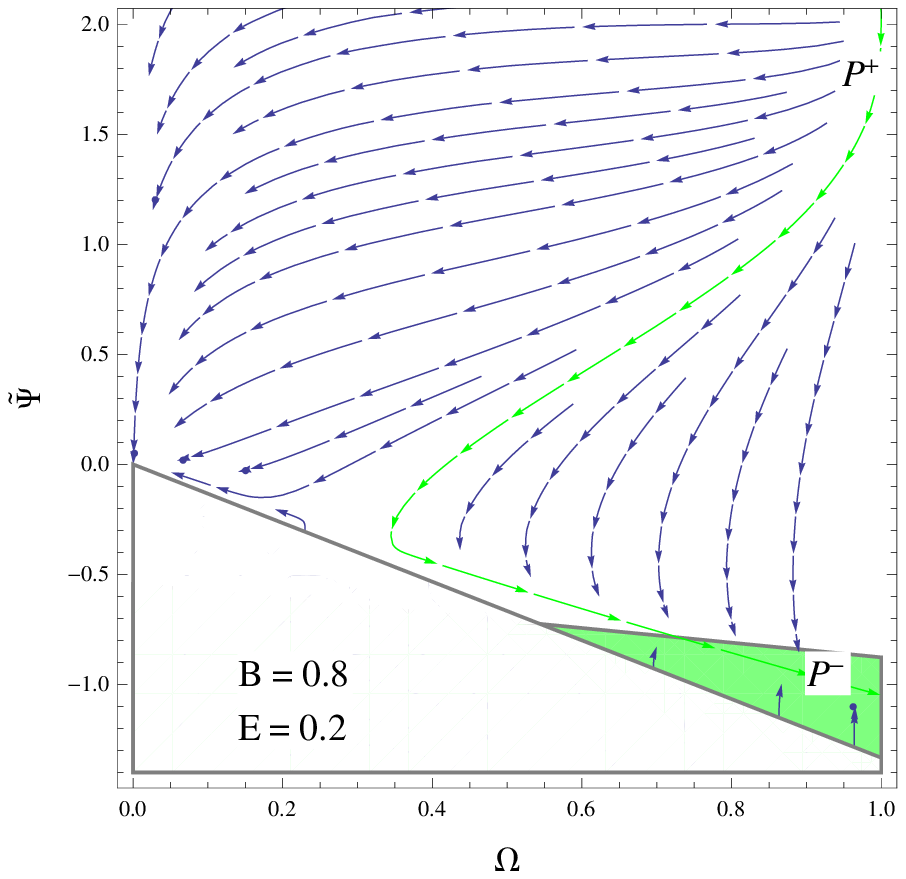,
width=0.4\linewidth}\caption{Plots for the phase plane evolution of
viscous radiating fluid with $\gamma=4/3$, $\nu=k=1$, $\zeta_{0}=1$,
$\alpha=0.01$ and different values of $B$ and $E$.}
\end{figure}

\subsubsection{Case II ($E=0$):}

It is well-known that NLED helps to diminish the initial singularity
in the early universe where only the primordial plasma identifies
matter \cite{23}. Some recent results indicate that a magnetic
universe is appropriate to avoid the initial singularity and
ultimately shows late time accelerated expansion \cite{24}.  Here we
assume the squared electric field $<E^2>$ to be zero such that the
magnetic field $(F=2B^2)$ rules over the universe known as
magnetized universe. Thus the energy density (\ref{1a}) and pressure
(\ref{1b}) take the form
\begin{eqnarray}\label{1c}
\rho_{B}&=&\frac{B^2}{2\mu_{0}}(1-8\mu_{0}\alpha B^2),
\\\label{1d}
p_{B}&=&\frac{B^2}{6\mu_{0}}(1-40\mu_{0}\alpha B^2).
\end{eqnarray}
The respective evolution plots are given in Figure \textbf{3}. For
$\nu=k=\sqrt{2/3}$ and $\zeta_{0}=1$, we find that sink lies in
green region showing the stability of accelerated expansion for the
magnetized universe. This region tends to decrease by increasing the
value of magnetic field $B$. The point $P^{0}_{d}$ behaves as saddle
for small values of magnetic field. It is mentioned here that
increasing values of bulk the viscosity and the parameters $\nu$ as
well as $k$ with different values of $B$ give rise to the stability
of accelerated expansion of the universe for different choices of
$B$. We also find that a smaller value of bulk viscosity show
decelerated expansion with increasing values of $B$.
\begin{figure}\center
\epsfig{file=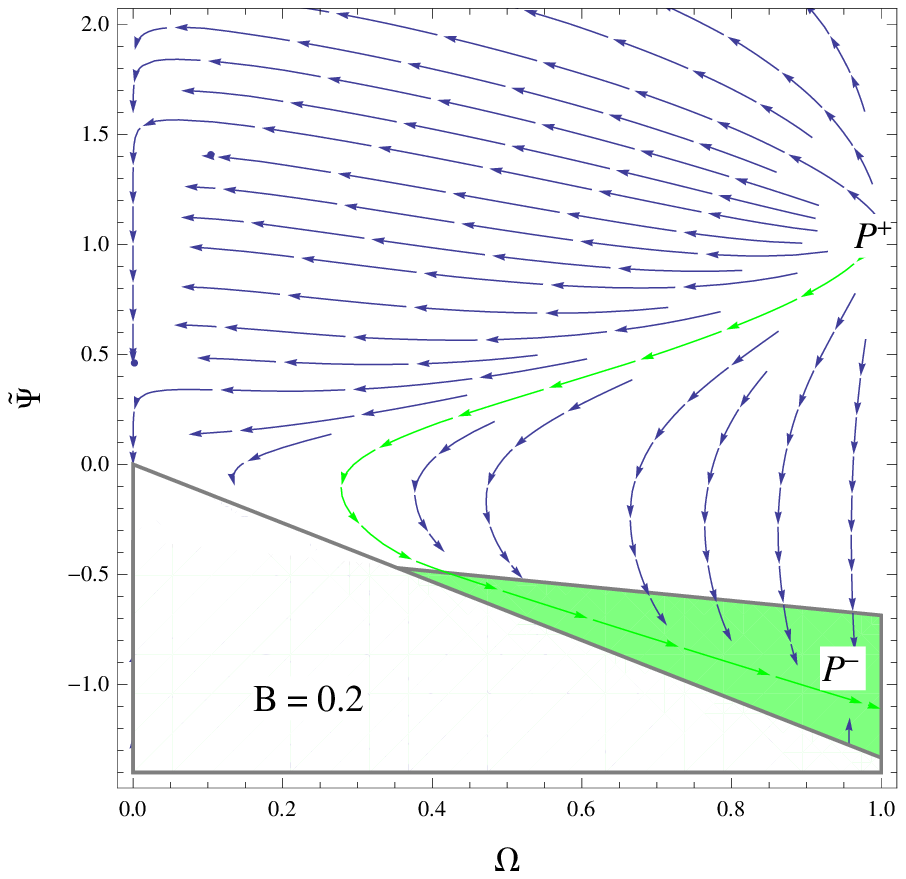,width=0.4\linewidth}\epsfig{file=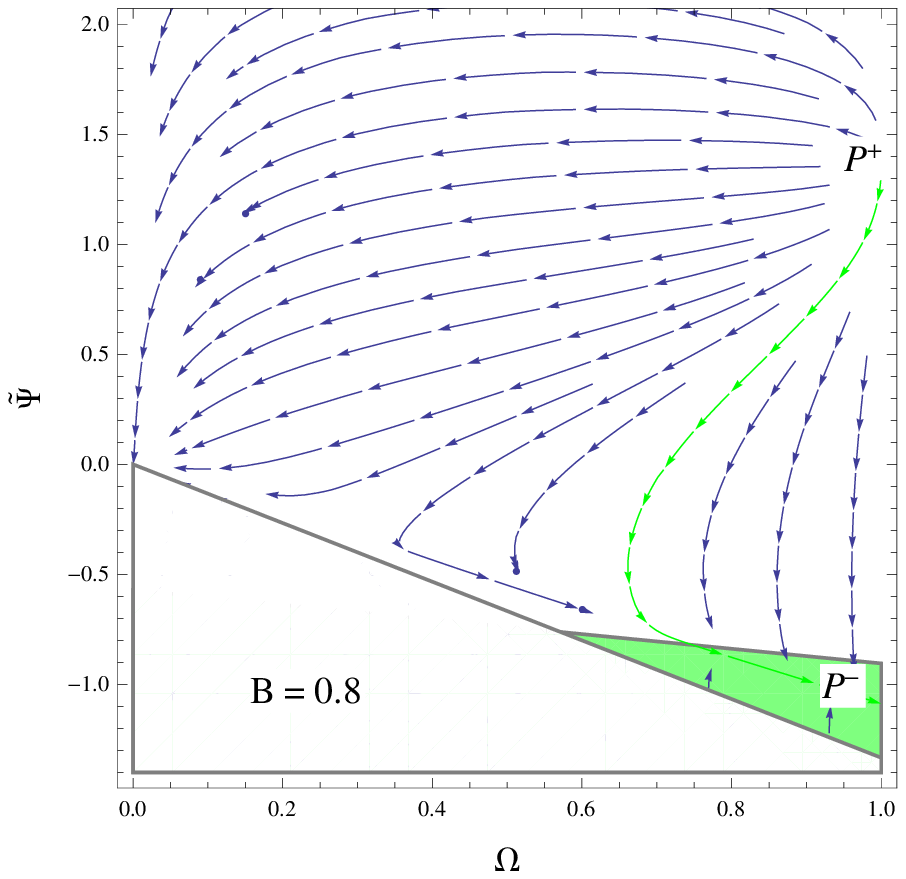,
width=0.4\linewidth}\caption{Plots for the phase plane evolution of
viscous radiating fluid with $\gamma=4/3$, $\nu=k=\sqrt{2/3}$,
$\zeta_{0}=1$, $\alpha=0.01$ and $B=0.2,0.8$.}
\end{figure}

\subsubsection{Case III ($B=0$):}

Here, we deal with the electric universe by setting $<B^2>=0$. The
corresponding energy density and pressure are given by
\begin{eqnarray}\label{2c}
\rho_{E}&=&\frac{E^2}{2\mu_{0}}(1+24\mu_{0}\alpha E^2),
\\\label{2d}
p_{E}&=&\frac{E^2}{6\mu_{0}}(1-8\mu_{0}\alpha E^2).
\end{eqnarray}
\begin{figure}\center
\epsfig{file=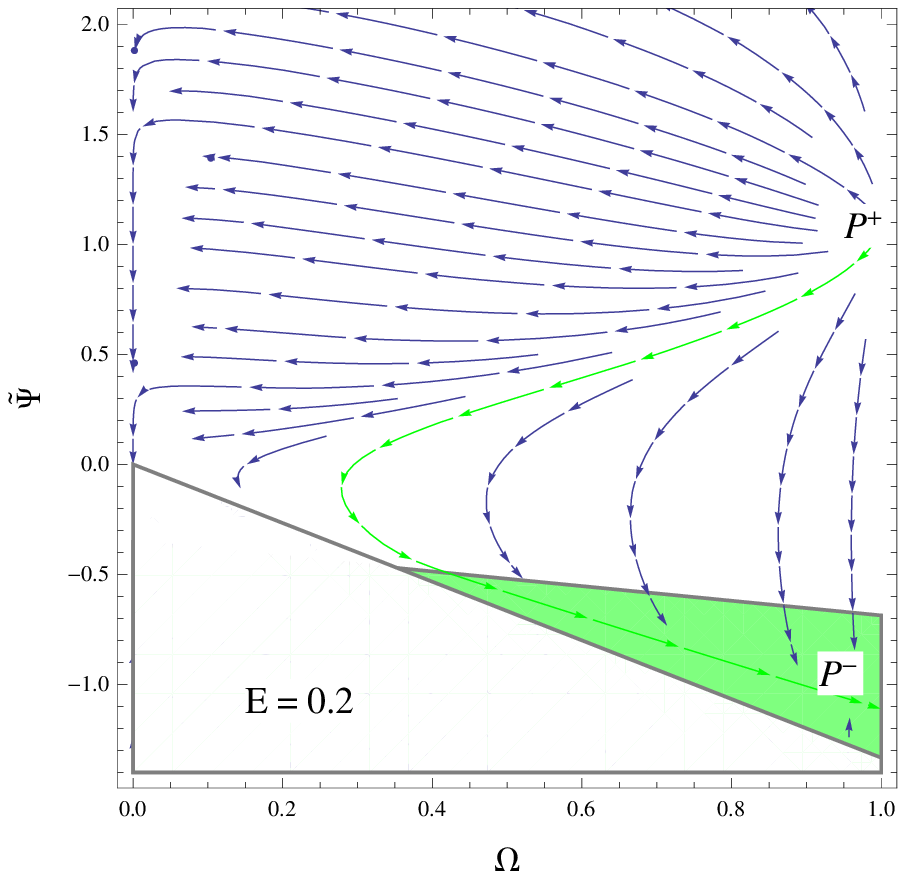,width=0.4\linewidth}\epsfig{file=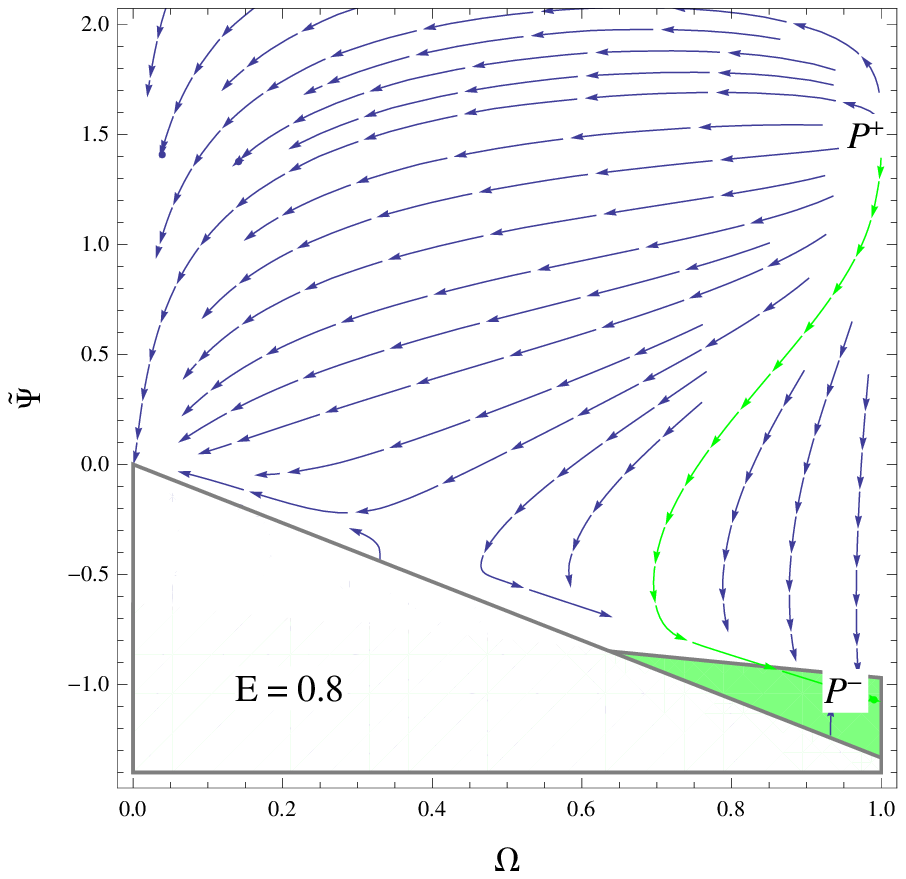,
width=0.4\linewidth}\caption{Plots for the phase plane evolution of
viscous radiating fluid with $\gamma=4/3$, $\nu=k=\sqrt{2/3}$,
$\zeta_{0}=1$, $\alpha=0.01$ and $E=0.2,0.8$.}
\end{figure}
The plots corresponding to different choices of electric field $E$
are shown in Figure \textbf{4}. For $\nu=k=\sqrt{2/3}$ and
$\zeta_{0}=1$, we analyze the sink $P^{-}_{d}$ in green region
showing accelerated expansion of the universe for different values
of $E$. The point $P^{0}_{d}$ behaves as a saddle for small values
of magnetic field. We find that the region for accelerated expansion
tend to decrease by increasing electric field $E$. It is observed
that accelerated expanding region exists for increasing values of
bulk viscosity and parameters $\nu$ as well as $k$ with all choices
of $E$. It supports the fact that the role of bulk viscosity and
electric field is to increase the stability of accelerated expansion
of the universe model. The summary of our results filled with
viscous radiating fluid is given in Table \textbf{1}.
\begin{table}[bht]
\textbf{Table 1:} \textbf{Stability Analysis of Critical Points for
Radiation Dominated Fluid}\\\\ \vspace{0.5cm} \centering
\begin{small}
\begin{tabular}{|c|c|c|c|}
\hline\textbf{Critical Point}&\textbf{Behavior}&
\textbf{Stability}\\
\hline{\textbf{$P_{r}^+$}}&{Source}&{Unstable}\\
\hline{\textbf{$P_{r}^-$}}&{Sink}&{Stable}\\
\hline{\textbf{$P_{r}^0$}}&{Saddle/Sink}&{Unstable/Stable}\\
\hline
\end{tabular}
\end{small}
\end{table}

\section{Power-Law Scale Factor}

In this section, we discuss the power-law behavior of scale factor
corresponding to the critical points. For this purpose, we integrate
Eq.(\ref{13}) which leads to
\begin{equation}\label{28}
\dot{\Theta}=-\frac{1}{2}\left[1+3p_{EM}
+(\gamma-1)\Omega+\tilde{\Psi}\right]\Theta^2.
\end{equation}
For $\Theta\neq0$, we formulate power-law scale factor whenever
$1+3p_{EM} +(\gamma-1)\Omega+\tilde{\Psi}\neq0$. Solving
$\Theta=\frac{3\dot{a}}{a}$ for $a(t)$, we obtain the generic
critical point as
\begin{equation}\label{29}
a=a_{0}(t-t_{0})^{\frac{2}{3[1+3p_{EM}+(\gamma-1)\Omega_{c}
+\tilde{\Psi_{c}}]}}.
\end{equation}
The following condition must hold for exponentially expanding models
(identified by the condition $1+3p_{EM}
+(\gamma-1)\Omega+\tilde{\Psi}=0$) to be present in the physical
phase space region (bounded by Eq.(\ref{18}))
\begin{equation}\label{30}
(1-\gamma)\Omega_{c}-3p_{EM}-1>-\frac{\gamma\nu^2}{k^2}\Omega.
\end{equation}
This condition is not satisfied in the physical phase space for
$\nu^2=k^2$. If $\nu^2>k^2$, the above inequality must be satisfied
in the following physical phase space region
\begin{equation}\label{31}
(1+3p_{EM})\left[1-\gamma\left(1-\frac{\nu^2}{k^2}\right)\right]^{-1}
<\Omega\leq 1.
\end{equation}
It is mentioned here that sign of the term $1+3p_{EM}
+(\gamma-1)\Omega+\tilde{\Psi}$ is quite important to evaluate
different cosmological stages. If $1+3p_{EM}
+(\gamma-1)\Omega+\tilde{\Psi}=0$, it corresponds to the exponential
expansion of the universe model. Also, $1+3p_{EM}
+(\gamma-1)\Omega+\tilde{\Psi}\gtrless0$ yields accelerated
expansion or contraction of the cosmological model, respectively. If
$\nu^2<k^2$, the possibility of having accelerated expansion will
narrow down. Figure \textbf{5} shows the physical phase space region
(excluding the white region with negative entropy production rate)
whereas yellow and dark gray regions correspond to accelerated and
exponential expansion of the universe model for $v^2>k^2$,
respectively. Table \textbf{2} provides the polynomial behavior of
power-law scale factor for different critical points with $1+3p_{EM}
+(\gamma-1)\Omega+\tilde{\Psi}\neq0$.
\begin{figure}\center
\epsfig{file=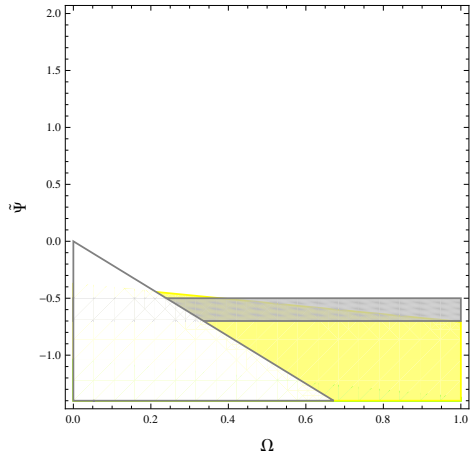,width=0.45\linewidth}\caption{Plot of
qualitative phase space analysis for power-law scale factor with
$v^2>k^2$. Yellow and dark gray regions indicate the accelerated and
exponential expansion of the universe model, respectively.}
\end{figure}
\begin{table}[bht]
\textbf{Table 2:} \textbf{Power-law Scale Factors for Different
Critical Points}\\\\ \vspace{0.5cm} \centering
\begin{small}
\begin{tabular}{|c|c|c|}
\hline\textbf{Critical Point}&\textbf{Scale factor for $\gamma=4/3$}\\
\hline{\textbf{$P_{r}^0$}}&{{$a_{0}(t-t_{0})^{-\frac{2}{9p_{EM}}}$}}\\
\hline{\textbf{$P_{r}^+$}}&{$a_{0}(t-t_{0})^{\frac{2}{3(3p_{EM}+
\tilde{\Psi}^{+}_{c}+\frac{4}{3})}}$}\\
\hline{\textbf{$P_{r}^-$}}&{$a_{0}(t-t_{0})^{\frac{2}{3(3p_{EM}+
\tilde{\Psi}^{-}_{c}+\frac{4}{3})}}$}\\
\hline
\end{tabular}
\end{small}
\end{table}

\section{Outlook}

In this paper, we have discussed the impact of NLED on the phase
space analysis of isotropic and homogeneous universe model by taking
noninteracting mixture of electromagnetic and viscous radiating
fluids. This analysis has been proved to be a remarkable technique
for the stability of dynamical system. An autonomous system of
equations has been developed by defining normalized dimensionless
variables. We have evaluated the corresponding critical points for
different values of the parameters to discuss stability of the
system. We have also calculated eigenvalues which characterize these
critical points. We summarize our results as follows.

Firstly, we have discussed stability of critical points through
their eigenvalues corresponding to different values of $E$ and $B$
for viscous radiation dominated universe model. It is found that the
critical points $P^{+}_{d}$ and $P^{-}_{d}$ correspond to source
(unstable) and sink (stable), respectively (Figures \textbf{1-2}).
It is mentioned here that the green region corresponds to
accelerated expansion of the universe. The point $P_{d}^-$ is a
global attractor in the physical phase space region which leads to
an expanding model dominated by viscous matter for various choices
of cosmological parameters. In the presence of both electric and
magnetic fields, we find that bulk viscosity increases the region
for accelerated expansion while the increasing values of $E$ shows
deceleration region for smaller values of bulk viscosity. It is
mentioned here that large values of bulk viscosity as well as other
parameters correspond to accelerated expansion of the respective
universe model for all choices of electric and magnetic fields.

We have also studied electric and magnetic universe cases
separately. It is found that sink lies in the green region showing
accelerated expansion of the magnetized universe for smaller values
of bulk viscosity and other parameters while increasing value of
magnetic field decreases this region. For $B=0$, we have analyzed
accelerated expansion of the universe model corresponding to large
values of the parameters which tends to decrease by increasing $E$.
It is worth mentioning here that the role of bulk viscosity is to
increase the green region for accelerated expansion with different
choices of $E$ and $B$ for both electric as well as magnetic
universe. Moreover, we have also studied the behavior of power-law
scale factor corresponding to the critical points. It is found that
the power-law scale factor indicates various phases of evolution
(accelerated or exponential expansion) of the respective universe
model.


\begin{thebibliography}{40}

\bibitem{1} Riess, A.G. et al.: Astron. J.
\textbf{116}(1998)1009; Perlmutter, S.J. et al.: Astrophys. J.
\textbf{517}(1999)565; Bennett, C.L. et al.: Astrophys. J. Suppl.
\textbf{148}(2003)1.

\bibitem{2} Sahni, V. and Starobinsky, A.A.: Int. J. Mod. Phys. A
\textbf{9}(2000)373; Tegmark, M. et al.: Phys. Rev. D
\textbf{69}(2004)03501.

\bibitem{5} Caldwell, R.R., Dave, R. and Steinhardt, P.J.: Phys. Rev.
Lett. \textbf{80}(1998)1582; Chiba, T., Okabe, T., Yamaguchi, M.:
Phys. Rev. D \textbf{62}(2000)023511.

\bibitem{6} Carroll, S.M., Hoffman, M. and Trodden, M.: Phys. Rev. D
\textbf{68}(2003)023509.

\bibitem{7} Gorini, V. et al.: Phys. Rev. D \textbf{69}(2004)123512.

\bibitem{8} Chimento, L.P.: Phys. Rev. D \textbf{69}(2004)123517.

\bibitem{9} Kamenshchik, A., Moschella, U. and Pasquier, V.: Phys. Lett. B
\textbf{511}(2001)265.

\bibitem{10} Bento, M. C., Bertolami, O. and Sen, A.A.: Phys. Rev. D
\textbf{66}(2002)043507.

\bibitem{10a} Heller, M., Klimek, Z. and Suszycki, L.: Astrophys. Space Sci.
\textbf{20}(1973)205; Zimdahl, W.: Phys. Rev. D
\textbf{53}(1996)5483.

\bibitem{10b} Vollick, D.N.: Phys. Rev. D \textbf{78}(2008)063524.

\bibitem{10c} Kruglov, S.I.: Int. J. Mod. Phys. D DOI:
10.1142/S0218271816400022.

\bibitem{10d} Ovgun, A.: arXiv: 1604.01837.

\bibitem{12} Copeland, E.J., Liddle, A.R. and Wands, D.: Phys. Rev. D
\textbf{57}(1998)4686.

\bibitem{13} Guo, Z.K. et al.: Phys. Lett. B \textbf{608}(2005)177.

\bibitem{13a} Garcia-Salcedo, R. et al.: arXiv: 0905.1103.

\bibitem{14} Yang, R.J. and Gao, X.T.: Class. Quantum Grav.
\textbf{28}(2011)065012.

\bibitem{15} Xiao, K. and Zhu, J.: Phys. Rev. D
\textbf{83}(2011)083501.

\bibitem{16} Acquaviva, G. and Beesham, A.: Phys. Rev. D
\textbf{90}(2014)023503.

\bibitem{aa} Garcia-Salcedo, R. and Bretonn, N.: Class.
Quantum Grav. \textbf{22}(2005)4783.

\bibitem{10bb} De Lorenci, V.A. et al.: Phys. Rev. D
\textbf{65}(2002)063501.

\bibitem{10bbb} Bandyopadhyay, T. and Debnath, U.: Phys. Lett. B
\textbf{704}(2011)95.

\bibitem{bb} Sharif, M. and Waheed, S.: Astrophys. Space Sci.
\textbf{346}(2013)583.

\bibitem{10b1} Novello, M. et al.: Class. Quantum Grav.
\textbf{24}(2007)3021.

\bibitem{10bb1} Zimdahl, W.: Phys. Rev. D \textbf{53}(1996)5483.

\bibitem{10b2} Novello, M. and Perez Bergliaffa, S.E.: Phys. Rep.
\textbf{463}(2008)127.

\bibitem{b} Maartens, R. and M\'{e}ndez, V.: Phys. Rev. D
\textbf{55}(1997)1937.

\bibitem{23} Giovannini, M., Shaposhnikov, M.: Phys. Rev. D
\textbf{57}(1998)2186.

\bibitem{24} De Lorenci, V.A. et al.: Phys. Rev. D
\textbf{65}(2002)063501; Novello, M., Bergliaffa, S.E.P. and Salim,
J.: Phys. Rev. D \textbf{69}(2004)127301; Novello, M. et al.: Class.
Quantum Grav. \textbf{24}(2007)3021.

\end{thebibliography}
\end{document}